\documentclass[twocolumn]{aastex62}

\usepackage{natbib}
\usepackage{hyperref}
\usepackage{amsmath}

\usepackage{longtable}
\usepackage{rotating}
\usepackage{booktabs}
\usepackage{comment}
\usepackage{xcolor}

\PassOptionsToPackage{hyphens}{url}
\hypersetup{
  colorlinks   = true, 
  urlcolor     = blue, 
  linkcolor    = blue, 
  citecolor   = blue 
}

\graphicspath{{./}}

\received{\today}
\revised{}
\accepted{}
\accepted{}

\shorttitle{Exomoons in the habitable zones of M dwarfs}
\shortauthors{H. Mart\'{i}nez-Rodr\'{i}guez et al.}

\begin{document} 

\title{Exomoons in the habitable zones of M dwarfs}

\author[0000-0002-1919-228X]{H\'{e}ctor Mart\'{i}nez-Rodr\'{i}guez}
\affiliation{Department of Physics and Astronomy and Pittsburgh Particle Physics, Astrophysics and Cosmology Center (PITT PACC), University of Pittsburgh, 3941 O'Hara Street, Pittsburgh, PA 15260, USA}
\correspondingauthor{H\'{e}ctor Mart\'{i}nez-Rodr\'{i}guez}
\email{hector.mr@pitt.edu}

\author[0000-0002-7349-1387]{Jos\'{e} Antonio Caballero}
\affiliation{Centro de Astrobiolog\'ia (CSIC-INTA), ESAC, Camino Bajo del Castillo, E-28691 Villanueva de la Ca\~nada, Madrid, Spain}

\author[0000-0003-1715-5087]{Carlos Cifuentes}
\affiliation{Centro de Astrobiolog\'ia (CSIC-INTA), ESAC, Camino Bajo del Castillo, E-28691 Villanueva de la Ca\~nada, Madrid, Spain}

\author[0000-0001-6806-0673]{Anthony L. Piro}
\affiliation{Carnegie Observatories, 813 Santa Barbara Street, Pasadena, CA 91101, USA}

\author{Rory Barnes}
\affiliation{Astronomy  Department, University  of  Washington, Box 951580, Seattle, WA 98195, USA}
\affiliation{NASA  Virtual  Planetary  Laboratory Lead Team, USA}

\begin{abstract}

M dwarfs host most of the exoplanets in the local Milky Way. 
Some of these planets, ranging from sub-Earths to super-Jupiters, orbit in their stars' habitable zones (HZs), although many likely possess surface environments that preclude habitability.
Moreover, exomoons around these planets could harbor life for long timescales and thus may also be targets for biosignature surveys. Here we investigate the potential habitability, stability, and detectability of exomoons around exoplanets orbiting M dwarfs.
We first compile an updated list of known M-dwarf exoplanet hosts, comprising 109 stars and 205 planets. 
For each M dwarf, we compute and update precise luminosities with the Virtual Observatory spectral energy distribution Analyzer and \textit{Gaia} DR2 parallaxes to determine inner and outer boundaries of their HZs.
For each planet, we retrieve (or, when necessary, homogeneously estimate) their masses and radii, calculate the long-term dynamical stability of hypothetical moons, and identify those planets that can support habitable moons.
We find that 33 exoplanet candidates are located in the habitable zones of their host stars and that four of them could host Moon- to Titan-mass exomoons for timescales longer than the Hubble time.

\end{abstract}





\section{Introduction} \label{sec:Introduction}

M dwarfs constitute only about one-third of the stellar mass of the Galaxy \citep{Chb03}, but about two-thirds of all its stars \citep{Gou96,RG97,Hen06,Boc10}.
Similarly to their more massive stellar counterparts, M dwarfs also host exoplanets \citep[among many others]{Del98, Mar98,But04,Bon05,Bon07,Bon13a,Jon07,Udr07,Dre15}.
Some of them orbit in the M-dwarf habitable zone (HZ), the circumstellar region where they can sustain liquid water on their surfaces \citep{Har79,Kas93,Sca07,Tar07,Ram18}.
The discoveries of such exoplanet candidates in temperate regions around their stars  \citep[e.g.][]{Udr07,Vog10,Vog12,Ang13,Ang16,Tuo13,Dit17,Gil17,Rei18} have come alongside with theoretical calculations of HZs \citep{Und03,Sel07,Kop13,Zso13,Kop14,Bar15,Joh18} and observational projects that aim to either discover Earth-like planets smaller than 1\,$M_\oplus$ \citep[e.g.][]{Nut08,Zec09,Bon13a,Art14,Qui14a,Ric15,Sul15} or characterize their atmospheres \citep{Wun19}. 

The M dwarfs' ubiquity and long main-sequence lifetime may favor their planets to host biological organisms \citep{Seg05,Loe16,Mul18}.
However, given the low luminosities of M dwarfs \citep{Del00,RH05,Ben16,Sch19}, Earth-like planets in their HZs are located at short orbital separation to their stars, which enhances the probability to detect transits or induced radial velocity (RV) signals, translates into synchronous rotation, and consequently into tidal locking \citep{Dol64,Pea77a,Pea77b,Jos97,Mur99,Tar07,Kit11,Bar17}.
This induced synchronous rotation has strong implications on the habitability of M-dwarf hosted planets because a hemisphere always faces the star and the other remains in complete darkness, introducing a spatially dependent incoming radiation and a temperature gradient between both hemispheres.
Models of atmospheric circulation in dry, tidally locked, rocky exoplanets have shown that this circulation acts as a global engine that constrains large-scale wind speeds \citep{Hen11,Sho11,Yan13,Kol16}.
Besides, in exoplanets that orbit close to their host stars, tidal interaction can affect climate stability, apart from spin and orbital properties \citep{Kas93,Bar08,Hel11a,Hel11b,Sho14}. 
In addition, flaring activity, which is frequent in M dwarfs \citep[and references therein]{Jef18,Loy18}, can cause strong atmospheric erosion in their exoplanets, with greater incidence on those that have no, or weak, magnetic moments \citep{Buc06,Cno07,Lam07,Seg10,Vid13,Sho14,Fra18,How18,Rod18}. On the other hand, \cite{DB15} showed that core convection could still drive a strong geodynamo for gigayears on tidally locked worlds.
Our knowledge of the atmospheres of Earth-like (and super-Earth) planets with one hemisphere under constant, strong irradiation is still incomplete, but the presence of a thick atmosphere or an ocean might help distribute the heat across the planet \citep[][see also \citealt{Pie11} for H$_2$-rich atmospheres]{Jos97,Jos03,Sca07,Tar07,Sel11,Hen12,Sho13,Kop14,Yan14,Wor15,Tur16,Way18}. 

One way to analyze the synchronous-rotation habitability problem of M-dwarf exoplanets is to introduce the presence of one or more hypothetical exomoons.
Contrary to their host planets, exomoons would not become tidally locked to the star, but more likely to the planet itself (\citealt{Hin13} -- see also \citealt{Gol66} and \citealt{Rey87} for spin-orbit coupling of moons in our solar system).
As a result, both exomoon hemispheres would receive the same amount of energy.
The presence of an exomoon would also alleviate a second habitability problem: except for a few cases, the best characterized M-dwarf hosted exoplanets are super-Earths or mini-Neptunes with masses greater than 4--5\,$M_\oplus$, which could be oceanic or gaseous worlds without any rocky surface, rather than Earth-like planets \citep{Leg04,Ray04,Lam09,Hil18}. 
In this picture, exomoons of about the mass of the Earth (with masses large enough to have a carbon--silicate cycle) would be habitable in the anthropocentric view, as opposed to their more massive planet hosts.
 	
Exomoons could also have effects on the habitability of their host exoplanets. For instance, the Moon contributes to steady the Earth's obliquity, which might have allowed life to develop in stable conditions for a sufficiently long time (\citealt{Las93,Las04,Lou01}; although see \citealt{Arm14,Dei18}). 
It is considered that moons are more stable and have longer lifetimes when the planet/moon systems are further from the parent star, the planets are heavier, or the parent stars are lighter \citep{Bar02,Sas12,Sas14}.
In addition to the flux received by the host star, a moon's climate and, thus, its habitability, can also be affected by planetary insolation and thermal emission, by periodic stellar eclipses, and by tidal heating \citep{Zol17}.
Tidal interactions between exoplanet and exomoon are a source of heating that is more relevant for the latter, such as in the inner Galilean moons \citep{Yod79,TP13,Mak14,Cla15,Tyl15,Ren18}.
Moderate tidal heating may help sustain tectonic activity, possibly a carbon--silicate cycle, and may also provide energy for biochemical reactions
\citep{Cam09,Kal10,Hel12,Hel13,Hin13,Hel14,For16,Zol17}. 

Exomoons are yet to be discovered (\citealt{Kip13,Kip15} -- but see the three {\em Kepler} candidates reported by \citealt{Sza13}, and especially \object{Kepler-1625\,b} -- \citealt{Tea18a, Tea18b,Hel19,Kre19}).
Most planet detection techniques are not applicable to that of exomoons due to the small relative sizes in comparison with their host planets.
One of the most relevant techniques is the transits method, but it relies upon favorable geometric conditions and  high photometric precision \citep{Kip09b,Kip12}. 
For the majority of exoplanet--exomoon phase curves, the lunar component is undetectable to the current technology of transits observations, and future surveys would require a photometric precision of a few parts per million \citep[][but see again \citealt{Tea18b}]{For17}.  
Nevertheless, the detectability of exomoons has been proposed feasible with current or near-future technology and a variety of methods:
microlensing \citep{Han02}, photocentric transit timing variation \citep{Sim07,Sim15}, modulation of planetary radio emissions \citep{Noy14}, spectroastrometry \citep{Ago15}, polarization of transits \citep{Ber18}, radial velocity \citep{Van18}, and transits of the plasma torus produced by a satellite (\citealt{BenJ14}; see the reviews by \citealt{Cab07}, \citealt{Per18}, and \citealt{Hel18}).
In particular, the European Space Agency (ESA) mission {\em CHaraterising ExOPlanet Satellite} ({\em CHEOPS}) could detect Earth-sized
exomoons with the transit method \citep{Sim15}.

Here, we explore the potential habitability of exomoons around known M-dwarf's exoplanets.
We compile a state-of-the-art list of M-dwarf exoplanetary systems, derive precise stellar luminosities from multiwavelength photometry and {\em Gaia} DR2 parallaxes, apply conservatively the radiative transfer equilibrium equation, identify exoplanets in HZ that can be orbited by exomoons, discuss on their detectability, and investigate their long-term dynamical stability based on the latest models by \cite{Pir18a}, who modeled exoplanets torqued by the combined tides of an exomoon and a host star in order to find timescales for the gravitational loss/destruction of the exomoon.

\section{Analysis} \label{sec:Analysis}

\subsection{Stellar and exoplanet sample} \label{subsec:Star_sample}

As of 2019 March 19, the Extrasolar Planets Encyclopaedia\footnote{\tt \url{http://exoplanet.eu/}} had reported a total of  4011 exoplanets with the label ``confirmed'' \citep{Sch11}.
In our analysis, firstly, we discarded the 258 planets detected by direct imaging (123), microlensing (88), pulsar timing (33), transit timing variation (8), and astrometry (6), 
many of which are located at large distances from the solar system and orbital separations, suffer from strong high-energy radiation, are difficult to characterize, or even are bona fide brown dwarfs \citep{Cab18}.

Of the remaining 3734 planets, 205 orbit 109 systems with M-dwarf hosts detected via transit and/or radial velocity. 
To identify them, we selected all planets orbiting stars with M spectral type in the Extrasolar Planets Encyclopaedia except for nine giants and subgiants (\object{HD~208527}, \object{HD~220074}), pre-main-sequence stars (\object{CVSO~30}, \object{V830~Tau}), and spectroscopic, eclipsing, and close binaries (\object{DW~UMa}, \object{HD~41004~B}, \object{Kepler-64}, \object{KIC~5095269~AB}, \object{NLTT~41135}).
Besides, we selected all planets orbiting stars without spectral type in the Encyclopaedia, but with masses lower than 0.71\,$M_\odot$, which is the largest mass of an M dwarf in the Encyclopaedia, and that have M spectral type in the NASA Exoplanet Archive\footnote{\tt \url{https://exoplanetarchive.ipac.caltech.edu/}}.
Among the 205 exoplanets, we also included the planets orbiting 
the trio of stars \object{BD--21~784}, \object{BD--06~1339}, and \object{Kepler-155} \citep{LoC13,Row14,Tuo14}, which have been spectroscopically classified as stars at the K--M boundary \citep{Upg72,Ste86,Zec09,Mui12a,Gaid13,Tuo14}, and \object{LP~834-042}\,c, which is listed in the Encyclopaedia as {\tt detection\_type = Other} but has a radial velocity confirmation \citep{Ast17b}.

The 109 M dwarfs and 205 exoplanets are listed in Tables~\ref{table:stars} and~\ref{table:planets}.	
For the stars, we tabulate common/discovery and alternative names\footnote{We follow this naming priority order: common (e.g. Proxima, Luyten's star), Bayer (e.g. $\alpha$~Cen~C), Flamsteed/variable (e.g. 3~Lyr, IL~Aqr, V1428~Aql), Lalande, Henry Draper, Bonner/C\'ordoba Durchmusterung, Wolf, Ross, Gliese-Jahreiss (but GJ number is always given), Luyten Palomar or equivalent, and discovery names (e.g. MCC, PM, 2MUCD, {\it Kepler}, {\it K2}, KOI, EPIC).}, 
equatorial coordinates 
and parallactic distances from {\em Gaia} DR2 \citep{Gaia18}, except for Luyten's star and Lalande~21185, for which we used {\em Hipparcos} parallactic distances \citep{van07}.
Distances range from merely 1.30\,pc (Proxima Centauri; \citealt{Ang16}) to about 430\,pc (\object{Kepler--235}; \citealt{Bor11}).
All stars further than 30\,pc have been discovered by {\em Kepler} (and {\em K2}) except for \object{NGTS--1} and \object{HATS--6} \citep{Har15,Bay18}.
Besides, most of them are field M dwarfs, but at least four belong to stellar cluster members or associations (namely \object{K2--25} in the Hyades --\citealt{Man16a}--, \object{K2--33} in Upper Scorpius --\citealt{Man16b}--, \object{K2--95} and \object{K2--264} in the Praesepe --\citealt{Pep17,Liv19}--).

For the planets, we tabulate their names, letters, main orbital parameters (semi-major axes $a$ and orbital periods $P_{\rm orb}$), and corresponding references.
For the 43 planets without a published semi-major axis \citep[e.g.,][]{Row14,Mui15,Sah16}, we derived them from their known periods and our estimated stellar masses, using Kepler's Third Law.
For the 89 planets discovered with the RV method and without measured transits, we tabulate their minimum masses, $M_2 \sin{i}$ (marked with `$>$'), while for the 103 planets discovered with the transit method and without RV follow-up, we tabulate their radii.
We have true masses and radii for 13 transiting planets with RV follow-up in 10 systems (see Table \ref{table:true}).

\begin{table}

\caption{Exoplanets with true masses and radii.}
\label{table:true}
\centering
\begin{tabular}{llll}
\hline
\hline
\noalign{\smallskip}
Star  & Planet & Transit & RV  \\
 	& 	 & References\tablenotemark{a} & References\tablenotemark{a}  \\
\noalign{\smallskip}
\hline
\noalign{\smallskip}
 HATS-6 & b & Har15 & Har15 \\
\noalign{\smallskip}
 PM J11302+0735  & b & Clo17  & Sar18  \\
\noalign{\smallskip}
  Kepler 45 & b &  Jon12 & Jon12  \\
\noalign{\smallskip}
  LHS 1140 & b & Dit17 & Dit17 \\
\noalign{\smallskip}
  & c & Men19 & Men19 \\
\noalign{\smallskip}
  LHS 3275 & b & Har13 & Cha09 \\
\noalign{\smallskip}
  LP 424--4 & b & Bon12 & Bon12 \\
\noalign{\smallskip}
  LTT 3758 & b & Sou17 & Bon18 \\
\noalign{\smallskip}
  PM J11293-0127 & b & Sin16  & Alm15 \\
\noalign{\smallskip}
  & c & Sin16  & Alm15 \\
\noalign{\smallskip}
  & d & Sin16  & Alm15  \\
\noalign{\smallskip}
  Ross 905 & b & Mac14 & Tri18 \\
\noalign{\smallskip}
  USco~J161014.7-191909 & b & Man16b & Man16b \\
  \noalign{\smallskip}
\hline
\end{tabular}
\tablenotetext{$a$}{
Alm15: 	\cite{Alm15}; 
Bon12: \cite{Bon12};
Bon18b: \cite{Bon18b}; 
Cha09: 	\cite{Cha09}; 
Clo17: 	\cite{Clo17}; 
Dit17: 	\cite{Dit17}; 
Har13: 	\cite{Har13}; 
Har15: 	\cite{Har15}; 
Jon12: 	\cite{Jon12}; 
Mac14: 	\cite{Mac14}; 
Man16b: \cite{Man16b}; 
Men19: 	\cite{Men19}; 
Sar18:  \cite{Sar18};
Sin16: 	\cite{Sin16}; 
Sou17: 	\cite{Sou17}; 
Tri18: 	\cite{Tri18}.}
\end{table}

\subsection{Planetary masses and radii} \label{subsec:MR}

Radial-velocity (RV) measurements only yield the minimum mass of an exoplanet, whereas transit measurements only yield the radius.
For our habitability analysis, we used planet masses and radii.
However, only 13 planets have both RV and transit measurements, i.e. true masses and radii (see above).
As a result, for the other 192 planets, we derived planetary radii and masses. Among the multiple mass-radius relations available in the literature for different planetary compositions \citep[e.g.,][]{For07,Sea07,Sot07,Lis11,Bar13,Wu13,Wei14,Bar15,Rog15,Wol16,Che17,Kan19}, we chose that of \citet{Che17}, as it covers a broad range of planetary sizes.

The largest planet transiting an M dwarf is \object{NGTS--1}\,b, with $R \sim$ 1.33\,$R_{\rm Jup}$ \citep{Bay18}. On the other side of the relation, the smallest planets transit the stars \object{Kepler~138}\,b, \object{LSPM~J1928+4437}\,( b, c, and d), \object{Kepler--125}\,c, and \object{2MUCD~12171} (now universally known as TRAPPIST-1; d and h), and have masses and radii in the intervals 0.10--0.39\,$M_{\oplus}$ and 0.52--0.78\,$R_{\oplus}$, close to the values of Mars (0.107\,$M_{\oplus}$, 0.343\,$R_{\oplus}$).
However, the distributions of masses and radii of the whole planet sample peak around 5--10\,$M_{\oplus}$ and 1--3\,$R_{\oplus}$, respectively.

\subsection{Stellar luminosities and masses} \label{subsec:Luminosities}

First of all, we compiled {\it Gaia} DR2 equatorial coordinates (in J2015.5 epoch and J2000.0 equinox) with VizieR \citep{Och00} for 107 of our stars.
For the other two, \object{Lalande~21185} and \object{Luyten's star}, without a {\it Gaia} DR2 entry, we retrieved the equatorial coordinates from Two Micron All Sky Survey (2MASS; \citealt{Skr06}), which were separated by less than two months from the J2000 epoch. 
Next, we collected multiband broadband photometry from the ultraviolet to the mid-infrared from the literature: 
{\it GALEX} DR5 \citep[FUV, NUV;][]{Bia11}, 
SDSS DR9 \citep[$ugri$;][]{Ahn12}, 
UCAC4 \citep[$BV$, $gri$,;][]{Zac13}, 
Pan-STARRS1~DR1 \citep[$gri$;][]{Ton12,Cha16}, 
Tycho-2 \citep[$B_T$, $V_T$;][]{Hog00}
APASS9 \citep[$BV$, $gri$;][]{Hen16}, 
2MASS \citep[$JHK_s$;][]{Skr06}, 
\textit{Gaia} DR2 \citep[$G_{BP} ~ G ~ G_{RP}$;][]{Gaia16,Gaia18}, 
{\it WISE} \citep[$W1W2W3W4$;][]{Cut12}, and 
allWISE \citep[$W1W2W3W4$;][]{Cut14}.

For that, we used the Upload X-Match tool of the interactive graphical viewer and editor for tabular data TOPCAT \citep{Och00} with a search radius of 5\,arcsec and the ``Best'' find mode option.
We made sure that the automatic cross-match was right by carrying out an inspection of each individual target and datum using the Aladin interactive sky atlas \citep{Bon00}, which provides simultaneous access to digitized images of the sky, astronomical catalogs, and databases. 
Eventually, for the 109 stars we compiled 1682 individual magnitudes along with their uncertainties.

We homogeneously derived the bolometric luminosity and effective temperature for each star with the compiled photometric data and the Virtual Observatory Spectral Energy Distribution Analyzer\footnote{\tt \url{http://svo2.cab.inta-csic.es/theory/vosa/}} \citep[VOSA;][]{Bay08}.
We selected the best fit between a grid of BT-Settl-CIFIST theoretical spectral models \citep{Bara15} and our empirical data under the constraints $\log{g}$ = 4.5, 2000\,K $< T_{\rm eff} <$ 4500\,K, and [Fe/H] = 0.0. 
The VOSA fits are acceptable in all cases, except for the extremely young star \object{K2-33} in the Upper Scorpius association \citep{Dav16, Man16b, Sch19} and the metal-poor star \object{K2--155} \citep{Die18a,Hir18}.
Further details on the luminosity calculation will be described by C. Cifuentes et~al. (in preparation).

The luminosities of the 109 planet-host stars are displayed in Table~\ref{table:stars}.
As expected, the most luminous stars are at the K--M boundary, with $L \sim$ 0.14--0.17\,$L_\odot$.
In particular, the most luminous stars, \object{Kepler--155} and \object{Kepler--252}, have VOSA-derived $T_{\rm eff}$ = 4300--4400\,K, which are actually warmer than the hottest known M dwarfs (thus, their spectral types should be revised).
On the other side, the least luminous stars, with $L = (5-15) \times\,10^{-4}$\,$L_\odot$, are the latest and coolest ones, which in our case are \object{2MUCD~12171} and \object{Proxima Centauri}, which have M8.0\,V and M5.5\,V spectral types, respectively.

\begin{figure}
\centering
\includegraphics[width=\hsize]{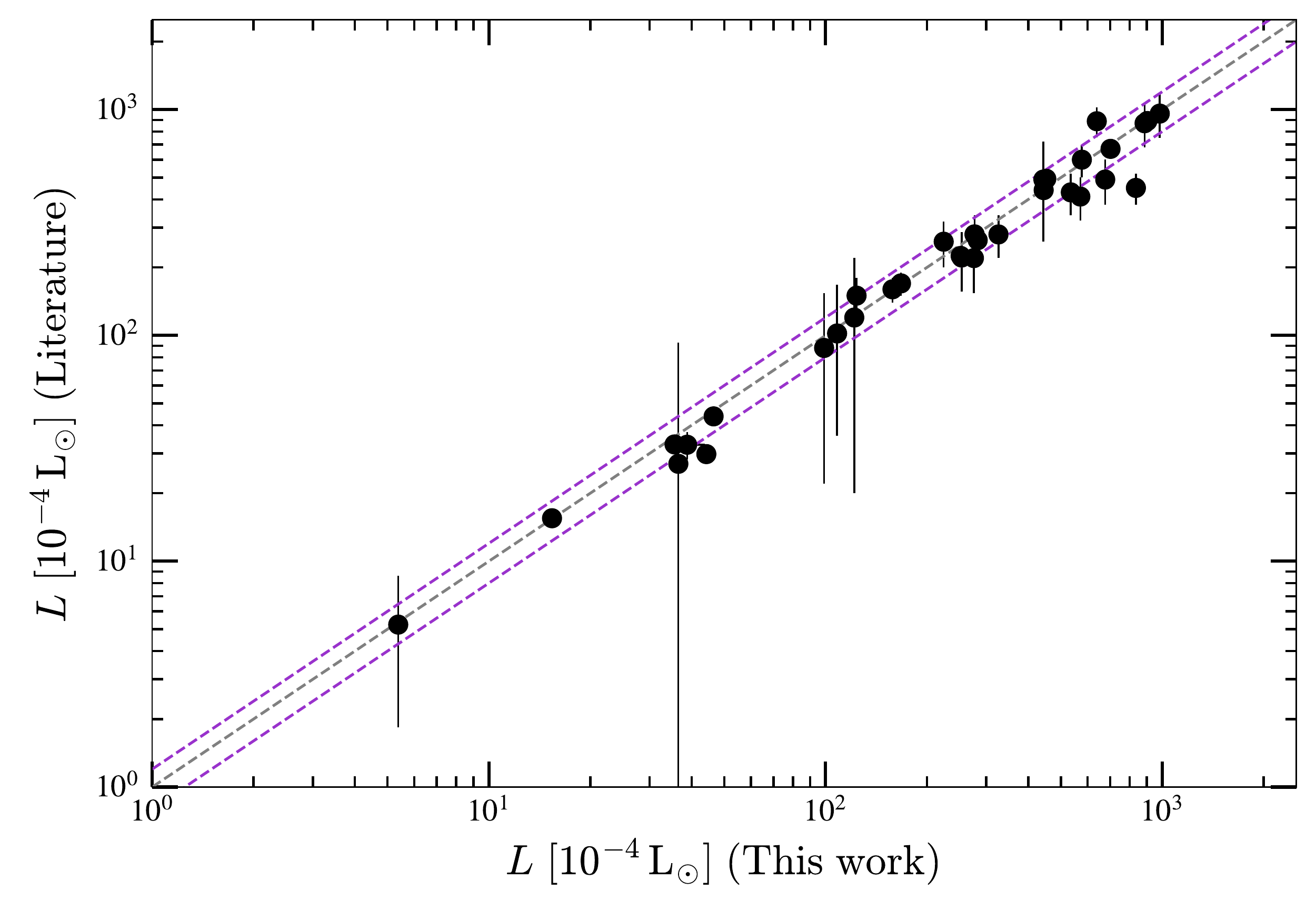}
\includegraphics[width=\hsize]{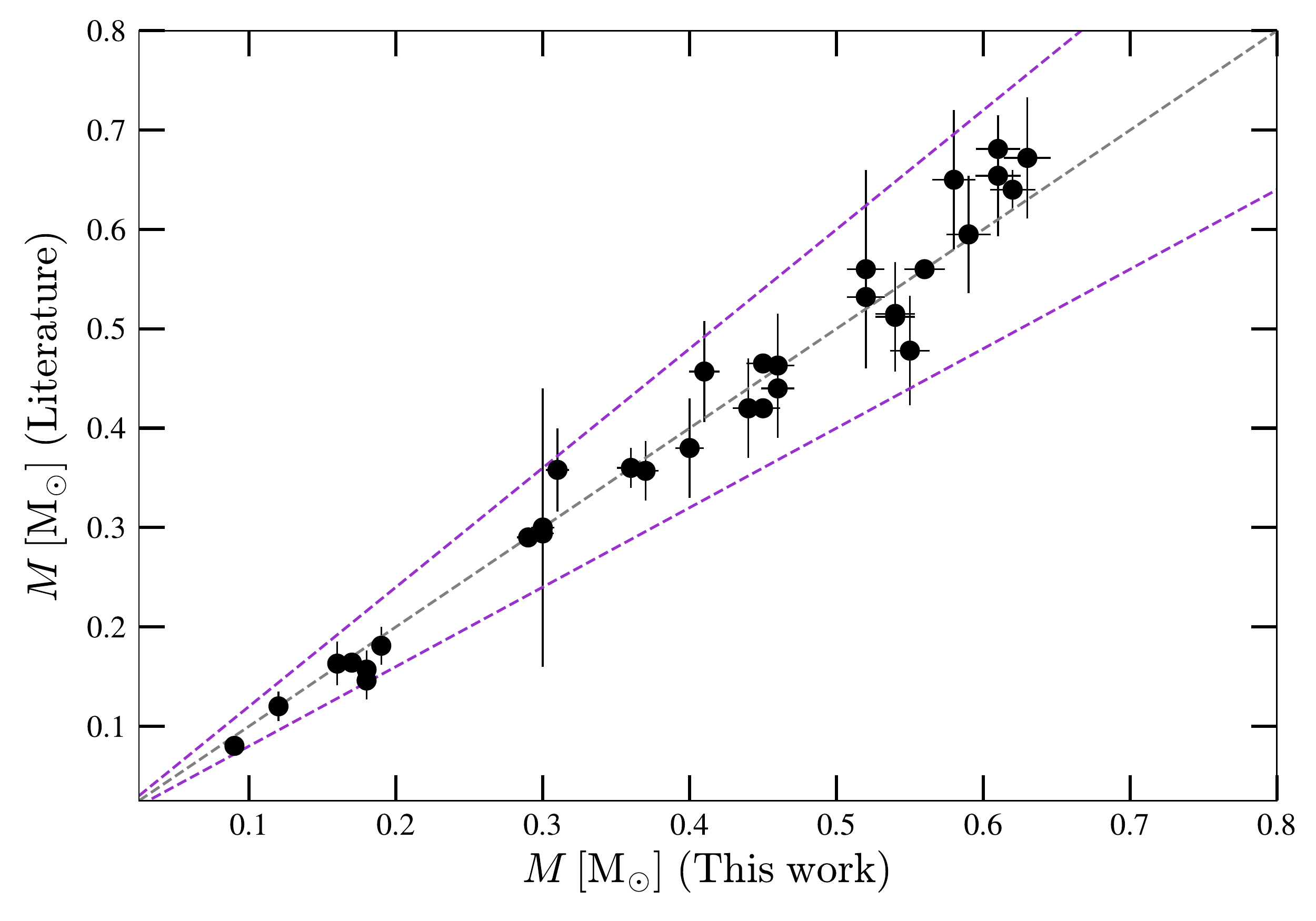}

\caption{Comparison between our luminosities (top) and masses (bottom) and those available in the literature.
The dashed lines indicate the 1:1 ratio and 20\,\% variations with respect to it.
}
\label{fig:Lcomparison}
\end{figure}

Finally, we computed stellar masses from $K_s$ magnitude and distance using the relations of \cite{Man19}.
We display them in the last column of Table~\ref{table:stars}.
The lowest masses, $M_*$ = 0.09--0.12\,$M_\odot$, are found again for \object{2MUCD~12171} and \object{Proxima Centauri}, and the highest ones, $M_*$ = 0.67--0.70\,$M_\odot$, correspond to the likely late-K dwarfs \object{Kepler--155} and \object{Kepler--252}, \object{K2--33} (actually incorrect, due to its extreme youth), and \object{K2--155} (probably due to a metallicity effect).
For comparison, we compiled published luminosities for 26 stars from the following works: \citet{Leg01,How10,Boy12,Gai14,Qui14,Har15,Sch16,Dit17,Gil17,Mal17,Bak18,Hir18,Rib18,Smi18,Aff19,Die19,Fei19,Gun19,Hob19,Per19}.
The luminosities derived in this work are in good agreement with the values given in the literature (see Figure \ref{fig:Lcomparison}).
Only five stars deviate more than 20\,\% of the 1:1 ratio, but only one deviates significantly from our measurement.
For \object{BD--17~400}, we calculated $L$ = (835$ \pm $13) $\times$ 10$^{-4}$\,L$_\odot$, but \citet{Leg01} tabulated $L$ = (450$ \pm $70) $\times$ 10$^{-4}$\,L$_\odot$ with pre-{\em Gaia} data.
The other four stars have luminosities consistent within error bars with our measurements (\object{LP~656--38}\footnote{\cite{Leg01} tabulated $L$ = (27$ \pm $66)\,10$^{-4}$\,L$_\odot$ for LP~656--38.}, \citealt{Leg01}; \object{GJ~96}, \citealt{Gai14}; \object{Kepler--186}, \citealt{Qui14}; \object{K2--149}, \citealt{Hir18}).

\section{Results and discussion} \label{sec:Results}

\subsection{Exoplanet habitability} \label{subsec:ExopHab}

We calculated an HZ for each host star using the one-dimensional climate models from \citet{Kop14}. We chose their ``recent Venus'' and ``maximum greenhouse'' estimates for the inner and outer HZ, respectively. The exoplanets' oceans might begin to evaporate and condense in the stratosphere at planetary equilibrium temperatures $\approx$ 320--340\,K \citep[``moist-greenhouse,''][]{Kas93,Kop13,Kop14}. However, between this process and the complete oceanic evaporation \citep[runaway greenhouse limit,][]{Kas93,Kop13,Kop14}, the exoplanets could still have surface temperatures adequate for habitability \citep[``continuosly habitable zone,''][]{Har78,Kas93}.

\begin{figure*}
\centering
\includegraphics[width=\hsize]{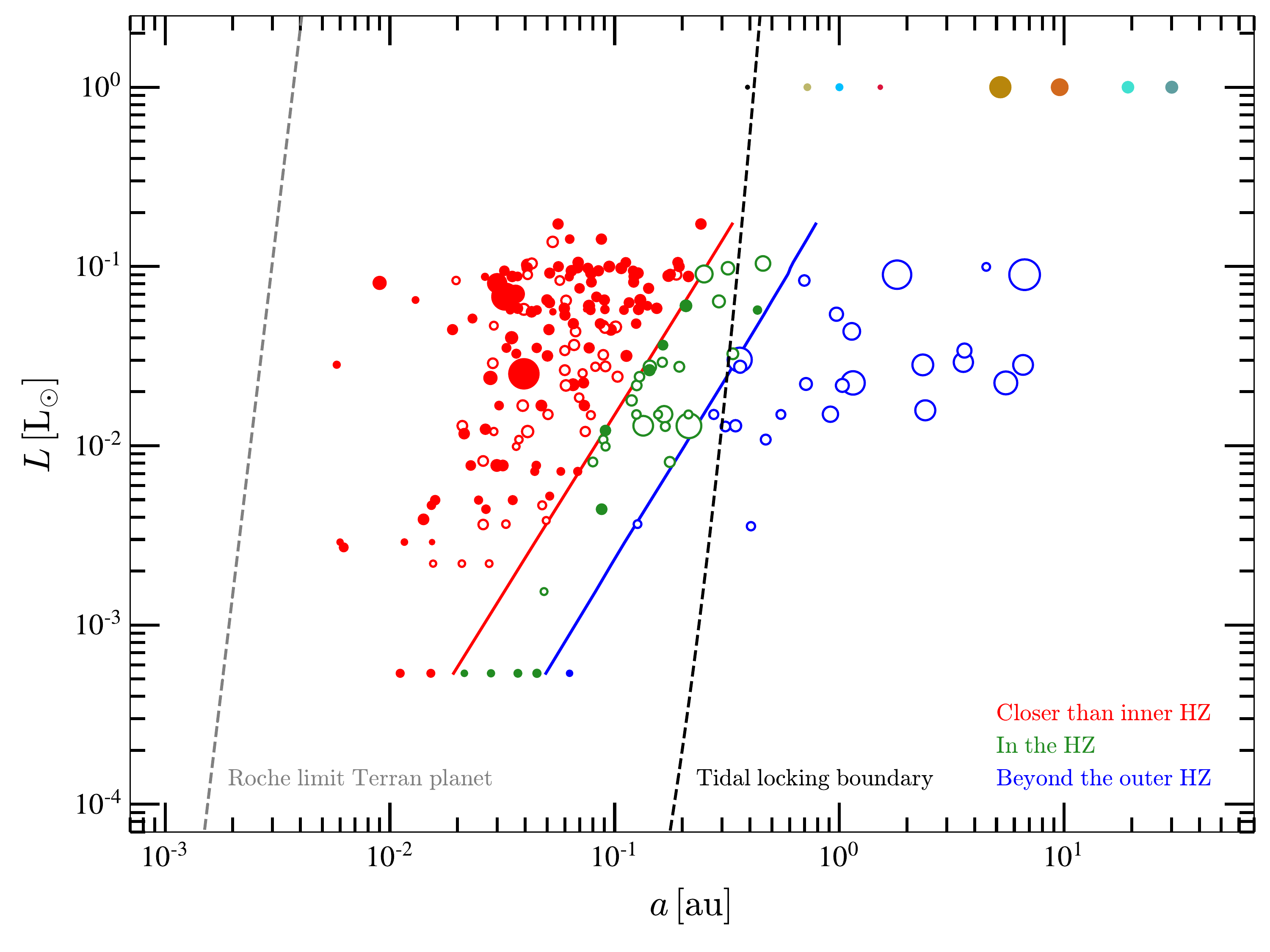}

\caption{Conservative HZ for all the M dwarfs in our sample (inner HZ, ``recent Venus,'' in red; outer HZ, ``maximum greenhouse,'' in blue). 
Their hosted exoplanets are depicted with sizes proportional to their masses in logarithmic scale, with filled circles if they have been detected by transit and with open circles if they have been detected by RV measurements. 
Exoplanets closer than the inner HZ, in the HZ, and beyond the outer HZ are shown in red, green, and blue, respectively. 
The dashed, gray line is the Roche limit for an Earth-like planet \citep{Agg74}. 
The dashed, black line represents a ``constant-time-lag'' tidal locking model for a 10 $M_{\oplus}$-planet with rapid initial rotation after 1\,Gyr \citep{Bar17}. 
The eight solar system planets are depicted in the upper part of the plot.}
\label{fig:HZ_extremes}
\end{figure*}

Figure \ref{fig:HZ_extremes} illustrates the HZ computation for our compiled exoplanet-host M dwarfs. The exoplanet candidates are depicted in three different colors, depending on whether they lie between their host stars and the inner HZ boundary (146), whether they lie between the inner and outer HZ boundary (33), or whether they lie beyond the outer HZ boundary (26). 
The 33 potentially habitable exoplanets are marked with the string ``Yes'' in column ``Potentially HZ'' from Table \ref{table:planets}. Of the 33 planet candidates, 10 have masses or minimum masses above 10\,$M_{\oplus}$ and 9 between 5 and 10\,$M_{\oplus}$.
Not counting the inclination angle effect in the RV planets, which translates into larger actual masses, most if not all of these 19 planets are mini-Neptunes and can hardly sustain liquid water on a rocky surface \citep{For07,Mil09,Dor17}.
In other words, these 19 planets lie within the HZ of their stars, but they are unlikely to be habitable.
However, their Earth-sized moons could actually be habitable.

As illustrated by Figure \ref{fig:HZ_extremes}, three habitable planets are beyond the approximate 1\,Gyr-tidal locking boundary: \object{BD--06~1339}\,c \citep{LoC13}, \object{Kepler--186}\,f \citep{Tor15}, and \object{HD~147379}\,b \citep{Rei18}.
Of them, two are Neptune-like planets and one, \object{Kepler-186}\,f, has a mass of only about 1.8\,$M_{\oplus}$.
However, following \citet{Bar17}, they will likely be tidally locked after 4.6\,Gyr, the solar system age (see Section \ref{subsec:Migration} for an individualized computation of planet locking times).

\subsection{Exomoons around Potentially Habitable Exoplanets} \label{subsec:Migration}

We focused on the 33 exoplanets orbiting the HZ of their host stars and analyzed whether or not they could host exomoons on reasonable time scales. 
For that, we followed the work of \citet{Pir18a}, who considered null exomoon eccentricity and obliquity for simplicity. 
If an exoplanet hosting an exomoon is close enough to its host star, the star's gravity  strips the exomoon away. 
This critical distance, namely the radius of the Hill sphere, is given by:

\begin{equation}\label{eq:acritm}
a_{\mathrm{crit},m} = f_{\rm crit} \left(\dfrac{M_{p}}{3M_{*}}\right)^{1/3},
\end{equation}

\noindent where $M_p$ and $M_{*}$ are the exoplanet and star masses, and $f_{\rm crit}$ = 0.49, a critical factor that assumes a prograde orbit \citep{Dom06,Pir18a}. 
Eventually, the exomoon will be stripped away by the host star if $a_{m} > a_{\mathrm{crit},m}$, or fall back into the planet if $a_{m} < a_{\mathrm{crit},m} \,$, where $a_{m}$ is the moon semimajor axis around the exoplanet.

If an exomoon gets too close to its host exoplanet, it will be tidally disrupted upon reaching the Roche limit, i.e., the point where the exomoon can no longer bind its material by gravitational forces. This condition is given by \citep{Pir18a}:

\begin{equation}\label{eq:aRoche}
a_{\rm{Roche}} \approx 1.34 \left(\dfrac{M_{p}}{\rho_{m}}\right)^{1/3},
\end{equation}

\noindent where $\rho_{m}$ is the exomoon's density \citep[see also][]{Agg74}. 
Under the assumption that all the initial angular momentum of an exoplanet eventually goes into its exomoon, the exomoon's semimajor axis can be written as:

\begin{equation}\label{eq:amP0}
a_{m} = \dfrac{4 \pi^{2} \lambda^{2} M_{p} R_{p}^{4}}{G M_{m}^{2} P_{0}^{2}},
\end{equation}

\noindent where $\lambda$ \citep[$\lambda_{\oplus} = 0.33$,][]{Wil94} is the exoplanet's radius of gyration \citep[$I_{p}/M_{p} R_{p}^{2}$, see][]{Pir18a}, $M_m$ is the exomoon mass, $I_p$ is the planet moment of inertia, $R_p$ is the planet radius, and $P_{0}$ is the planet initial spin period.
To estimate the exomoon migration timescale, namely, the e-folding timescale for the change in the exoplanet--exomoon distance, we used Eq.~25 from \citet{Pir18a}, which can be rewritten as:

\begin{equation}\label{eq:taumig}
\tau_{\mathrm{mig},m} = t_{k}  \left(\dfrac{k_{2, \oplus}}{k_{2}}\right) \left(\dfrac{\tau_{\rm{lag,\oplus}}}{\tau_{\rm{lag}}}\right) \left(\dfrac{a_m^{\, 8}}{M_m \, R_p^{\, 5}}\right),
\end{equation}

\begin{figure}
\centering
\includegraphics[width=\hsize]{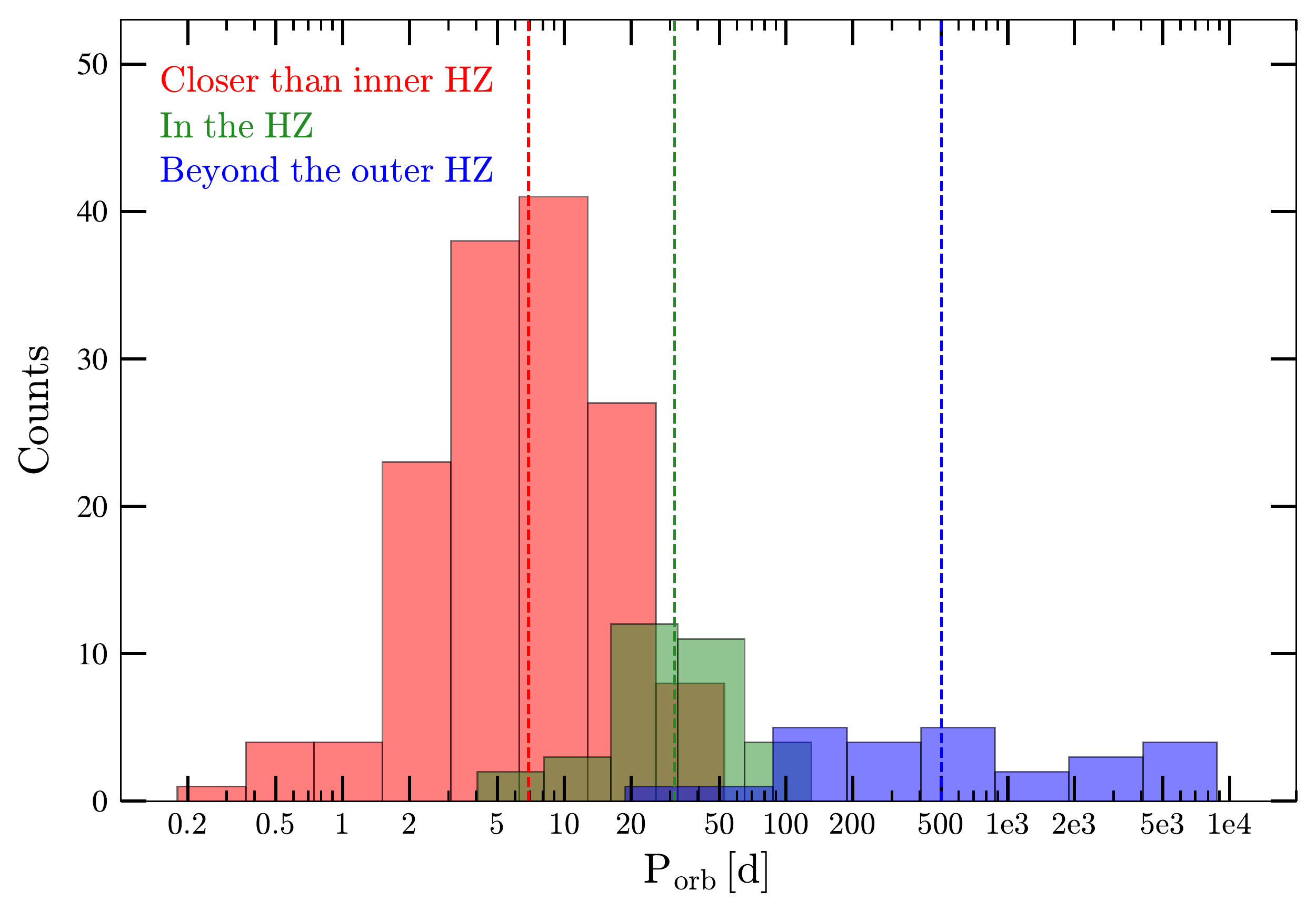}

\caption{Orbital periods for the exoplanets in our sample. 
The colors are the same as those in Figure~\ref{fig:HZ_extremes}.
The dashed, vertical lines show the median period for each subgroup.}
\label{fig:Porb}
\end{figure}

\noindent where $t_{k}$ = 250\,Gyr, and $a_m$, $M_m$, and $R_p$ are the exoplanet--exomoon separation, exomoon mass, and planetary radius in units of Earth--Moon separation, Moon mass, and Earth radius, respectively\footnote{a$_{\rm Moon}$ = 3.84399 $\times$ 10$^8$\,m, $M_{\rm Moon}$ = 7.346 $\times$ 10$^{22}$\,kg, and $R_{\oplus}$ = 6.3781366 $\times$ 10$^6$\,m.}. Here, $k_{2}$ is a Love number of degree 2 ($k_{2, \oplus} = 0.3$) and $\tau_{\rm{lag}}$ is the time lag between the passage of the perturber and the tidal bulge \citep{Bar17,Gre17,Pir18a}. 
For the Earth, $\tau_{\rm{lag},\oplus} = 638$ s \citep{Lam77,Ner97}.
The time lag $\tau_{\rm{lag}}$ is related with the tidal quality factor $Q$ via \citep{Hen09,Mat10,Bar13}:

\begin{equation}\label{eq:q}
Q^{-1} \approx n \tau_{\rm{lag}},
\end{equation}

\noindent where $n$ is the mean motion, which implies that

\begin{equation}\label{eq:taulag}
\dfrac{\tau_{\rm{lag,\oplus}}}{\tau_{\rm{lag}}} = \dfrac{Q}{Q_{\oplus}},
\end{equation}

\noindent where $Q_{\oplus} = 12$ \citep{Wil78}. Therefore, combining Eqs. \ref{eq:amP0}, \ref{eq:taumig} and \ref{eq:taulag},

\begin{equation}\label{eq:taumigapprox}
\tau_{{\rm mig},m} \, \propto \, Q \dfrac{R_p^{27} M_p^8}{P_0^{16} M_m^{17}}.
\end{equation}

Table \ref{table:taulag} summarizes the used values of $k_2$, $Q$ and $\lambda$ for four representative planetary types.
These parameters are reasonable extrapolations of $k_2$, $Q$, and $\lambda$ from models of exoplanets and actual measurements of planets of different sizes and compositions \citep{Dic94,Yod95,Aks01,Zha06,Hen09}.
For planets with sizes between $R_{\rm p}$ = 1.50\,$R_\oplus$ (small, rocky worlds) and $R_{\rm p}$ = 3.88\,$R_\oplus$ (Neptune), we fitted a linear function to $k_2$:

\begin{equation}\label{eq:k2r}
k_{2} = k_{2,0} + \gamma ~ R_{\rm p}.
\end{equation}

\noindent For $Q$ and $\lambda$, we fitted, respectively, two power laws and two linear functions, between $R_{\rm p}$ = 1.50\,$R_\oplus$ and $R_{\rm p}$ = 3.88\,$R_\oplus$, and between $R_{\rm p}$ = 3.88\,$R_\oplus$ and $R_{\rm p}$ = 10.97\,$R_\oplus$ (Jupiter):

\begin{equation}\label{eq:Qr}
Q = Q_{0} ~ R_{\rm p} ~ ^{\delta}
\end{equation}

\noindent and

\begin{equation}\label{eq:lambdar}
\lambda = \lambda_{0} + \epsilon ~ R_{\rm p}.
\end{equation}

The values resulting from these fittings are listed in Table \ref{table:k2Qlambda}. Next, we computed the parameters $a_{\rm Roche}$, $a_m$, and $\tau_{\mathrm{mig},m}$ for the 33 exoplanets in HZ with a hypothetical exomoon.
In our calculations, we set $M_{m} = M_{\mathrm{Moon}}$ and $\rho_{m}$ = 2500\,kg\,m$^{-3}$, which is a density intermediate between those of Io and Ganymede, but slightly lower than that of the Moon (that has a small iron core formed after the collision between Theia and the proto-Earth). 
The Roche limit, $a_{\rm Roche}$, smoothly depends on $\rho_m$ (Eq. \ref{eq:aRoche}).
For example, assuming the Earth density $\rho_{\oplus}$ = 5513\,kg\,m$^{-3}$, $a_{\rm Roche}$ would only be increased by 23\,\%.
Next, we set $a_{m} = a_{\mathrm{crit},m}$ in Eq.~\ref{eq:taumig} when the exomoon is stripped ($a_{m} > a_{\mathrm{crit},m}$) and used the value given by Eq.~\ref{eq:amP0} when it falls back ($a_{m} < a_{\mathrm{crit},m}$). We did not calculate $\tau_{\mathrm{mig},m}$ if $a_{m} < a_{\rm{Roche}}$. We remark that, for multiplanetary systems, these estimates may not be entirely accurate because of the impact of the other planets.

\begin{table}[]

\caption{Love number, tidal quality factor, and radius of gyration for different planetary types.
}
\label{table:taulag}
\centering
\begin{tabular}{lccc}
\hline
\hline
\noalign{\smallskip}
	Planet & $k_2$ & $Q$ & $\lambda$ \\ 
\noalign{\smallskip}
	type &  &  \\ 
\noalign{\smallskip}
\hline
\noalign{\smallskip}
  	Dry telluric  & 0.5  & 100 & 0.370 \\ 
\noalign{\smallskip}
 	Terran & 0.3  & 12 & 0.330 \\ 
\noalign{\smallskip}
 	Icy neptunian & 1.5  & $10^{4}$ & 0.230 \\ 
\noalign{\smallskip}
 	Gaseous jovian & 1.5  & $10^{6}$ & 0.254 \\ 
\noalign{\smallskip}
\hline
\end{tabular}
%
\end{table}

The initial planetary spin $P_{0}$ is one of the greatest unknowns in such estimates.
It is the orbital period, as opposed to the initial rotation period, which would be set by the last major impact.
Figure~\ref{fig:Porb} shows the orbital periods for all the exoplanets in our sample.
Minimum and maximum $P_{\rm orb}$ range from 4.3\,h \citep[\object{K2--137}\,b;][]{Smi18} to 24\,yr \citep[\object{BD--05~5715}\,c;][]{Mon14}, but they concentrate between 1 and 50\,days.
The range of $P_{\rm orb}$ for the exoplanets in HZ is narrower, from 4 to 130\,days, with a median value of 31\,days.
For completeness, the median $P_{\rm orb}$ for inner and outer HZ planets are 7\,days and 502\,days, respectively.

\begin{figure}
\centering
\includegraphics[width=\hsize]{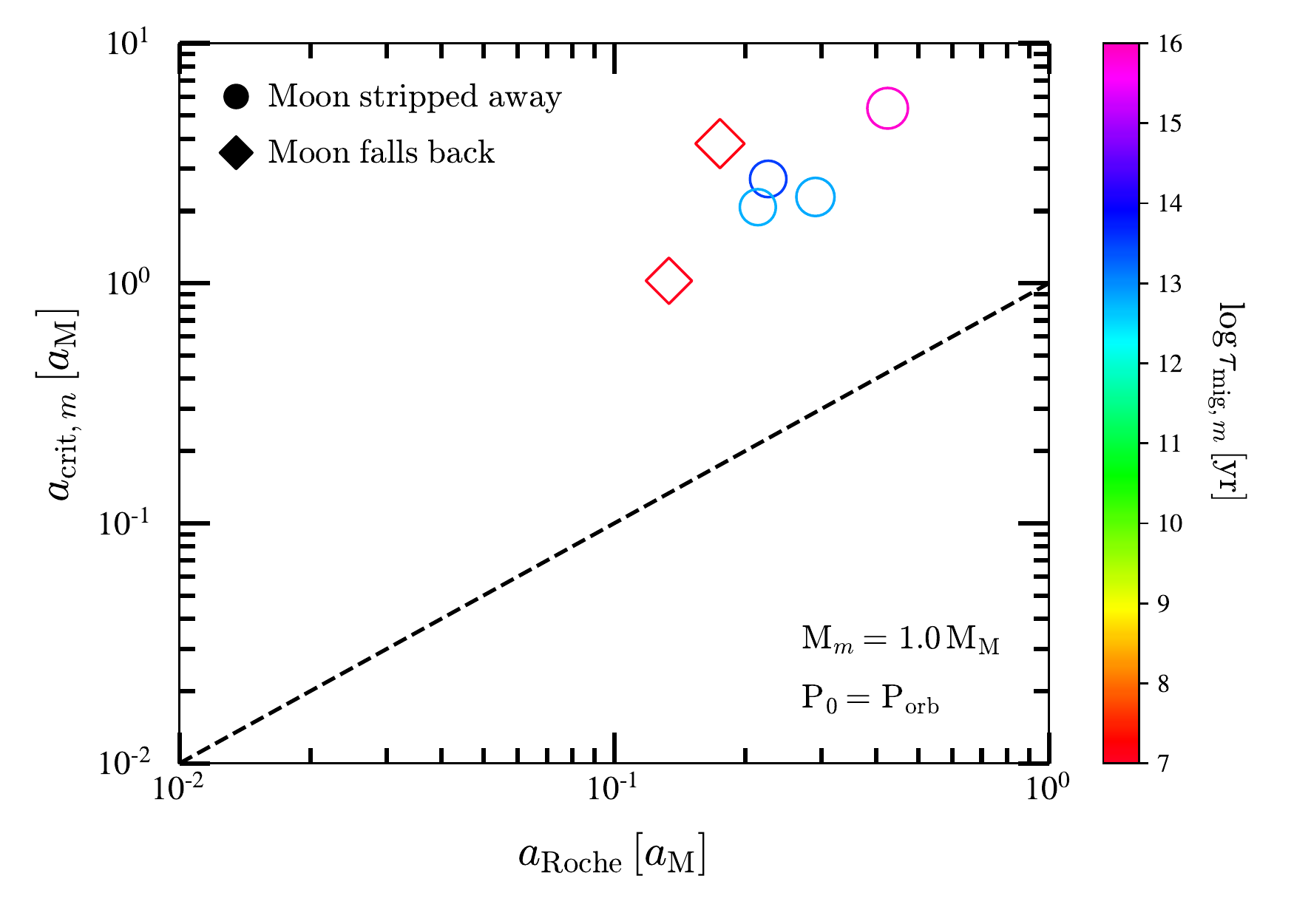}
\includegraphics[width=\hsize]{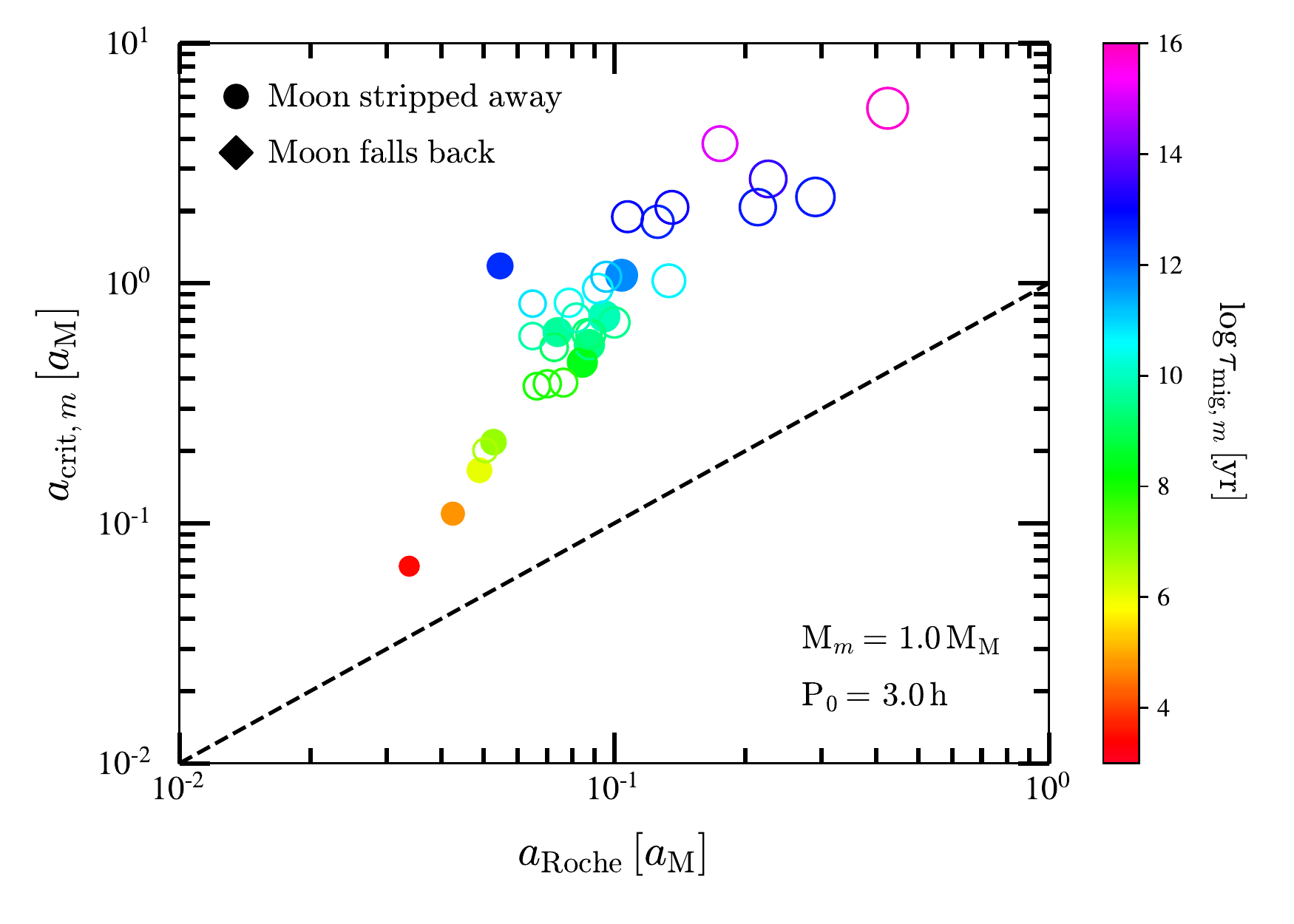}
\caption{$a_{\mathrm{crit},m}$ vs. $a_{\rm{Roche}}$ for the modeled exomoons around all the potentially habitable exoplanets in our sample. The circles and the squares (filled for primary transit exoplanet detection, empty for RV exoplanet detection) represent an exomoon being stripped away by the host star or falling back into the exoplanet, respectively. They are depicted with sizes proportional to their exoplanet hosts' masses in logarithmic scale. The black, dashed line corresponds to the exomoon tidal disruption limit, so no exomoon can exist below it. Top: $P_{0} = P_{\rm{orb}}$ for each planet. Bottom: $P_{0} = 3.0$ h.}
\label{fig:Migration}
\end{figure}

Figure~\ref{fig:Migration} depicts $a_{\mathrm{crit},m}$ vs. $a_{\rm{Roche}}$ for the 33 potentially habitable planets in our sample. 
It shows the two most extreme cases:  $P_{0,1} = P_{\rm{orb}}$ and  $P_{0,2}$ = 3.0\,h.
On the one hand, $P_{0,1} = P_{\rm{orb}}$ ($\tau_{{\rm mig},m} \, \propto \, Q\ R_p^{27} M_p^8 P_0^{-16}$) is the longest allowed value of $P_0$ because the exoplanet is tidally locked to its star. 
It is probably the most realistic case, as the habitable exoplanets in our sample orbit their stars under synchronous rotation. 
In particular, we computed the tidal locking time $t_{\rm lock}$ with ``constant-phase lag'' (CPL, where the phase between the perturber and the tidal bulge is constant and insensitive to orbital and rotational frequencies) and ``constant-time lag" (CTL, where the time interval between the perturber's passage and the tidal bulge is constant) models, using the publicly available \texttt{EqTide}\footnote{\tt \url{https://github.com/RoryBarnes/EqTide}} code and the \object{Proxima Centauri}\,b system as a template \citep{Bar16,Bar17}. We assumed zero eccentricity and obliquity, an initial planetary spin period of 3.0 h, a stellar Love number $k_{2,*} = 1.5$, a stellar gyration radius $\lambda_{*} = 0.1$ \citep[$\lambda_{\odot} = 0.061$,][]{Bou13}, and the planetary parameters $k_{2}, Q, \lambda$ varying accordingly with $R_{p}$. We chose the minimum $t_{\rm lock}$ between the calculations from the CPL and the CTL models.

%

\begin{table}[]
\small

\caption{Parameters for deriving $k_2$ (Eq.~\ref{eq:k2r}), $Q$ (Eq.~\ref{eq:Qr}), and $\lambda$ (Eq.~\ref{eq:lambdar}).}
\label{table:k2Qlambda}
\centering
\begin{tabular}{ccccccc}
\hline
\hline
\noalign{\smallskip}
	$R_{\rm{p}}$ & $k_{2,0}$ & $\gamma$ & $Q_{0}$ & $\delta$ & $\lambda_{0}$ & $\epsilon$  \\
\noalign{\smallskip}
	  &   & [$R_{\oplus}^{-1}$] & [$R_{\oplus}^{-\delta}$] &  &  &  [$R_{\oplus}^{-1}$] \\
\noalign{\smallskip}
\hline
\noalign{\smallskip}
  	$\leq$1.50  & 0.3 & 0  & $10^{2}$ & 0 & 0.37 & 0 \\
\noalign{\smallskip}
 	1.50--3.88 & --0.456 & 0.504  & 14.019 & 4.846 & 0.458 & --0.059 \\
\noalign{\smallskip}
 	3.88--10.97 & 1.5 & 0 & 24.600 & 4.431 & 0.217 & 0.003 \\ 
\noalign{\smallskip}
 	10.97--26 & 1.5 & 0 & $10^{6}$ & 0 & 0.254 & 0 \\ 
\noalign{\smallskip}
\hline
\end{tabular}
\end{table}

We found that only four habitable planets have $t_{\rm lock}$ between 0.1\,Gyr and 1\,Gyr, namely
\object{LP~834--042}\,d \citep{Ast17b},
\object{PM~J11293-0127}\,d \citep{Alm15,Sin16},
\object{HD~156384~C}\,e \citep{Ang13}, and
\object{CD--44~11909}\,c \citep{Tuo14},
and 9 greater than 1\,Gyr,
\object{GJ~96}\,b \citep{Hob18},
\object{CD--23~1056}\,b \citep{For11},
\object{BD--06~1339}\,c \citep{LoC13},
\object{Ross~1003}\,b \citep{Tri18},
\object{HD~147379}\,b \citep{Rei18},
\object{V1428~Aql}\,b \citep{Kam18},
\object{Kepler--186}\,f \citep{Tor15}, and
\object{IL~Aqr}\,b and~c \citep{Tri18}. 
In this situation, all of them could host exomoons beyond the Roche limit after all the material in the protoplanetary disk is dissipated \citep[10$-$30 Myr,][]{Ber89,Har98,Cal02}.
All other habitable planets get tidally locked in less than 90\,Myr.

However, the moon migration timescales are too low. 
In this scenario, only for six habitable exoplanets the condition $a_{m} > a_{\rm Roche}$ holds. 
The migration timescales are greater than $t_{\rm H}$, being $t_{\rm H} =$ 14.4\,Gyr the Hubble time (which is slightly longer than the universe age at 13.8\,Gyr), for four exoplanets (\object{CD--23~1056}\,b, \citealt{For11}; \object{Ross 1003}\,b, \citealt{Hag10,Tri18}; \object{IL~Aqr}\,b and c, \citealt{Tri18}), whose properties are shown in the top part of Table~\ref{table:Migration} and discussed in Section~\ref{subsec:HabMoons}.
The four of them have 
$a_{m} > a_{{\rm crit}, m} > a_{\rm Roche}$,
the largest planetary radii in the HZ ($R_p = $ 12--14$\,R_\oplus$), 
and $\tau_{\rm mig} \gg $ 0.8\,Gyr, which was the Earth's Late Heavy Bombardment duration \citep{Fas13,Nor14}. 
In this $P_{0,1} = P_{\rm{orb}}$ scenario, the hypothetical moons around the other 29 planets would fall onto them on time scales much shorter than protoplanetary disk dissipation times under the \texttt{EqTide} layout, not counting potential viscosity in the protomoon disk \citep{Lin86,Pol96,Mon06,Sas10}. 
The fifth and sixth habitable exoplanets for which $a_{m} > a_{\rm Roche}$, \object{LP~834--042}\,b 
and \object{BD--06~1339}\,c, have short $\tau_{\rm mig}$ of about 50 and 18\,Myr, respectively, which are comparable to the typical protoplanetary disk dissipation time \citep{Ida04,Her07,Lam12}.

On the other hand, $P_{0,2}$ = 3.0\,h ($\tau_{{\rm mig},m} \, \propto \, Q\ R_p^{27} M_p^8$) is the smallest value used by \cite{Pir18a}, similar to the shortest $P_{\rm{orb}}$ in our sample ($P_{0,2} \approx$ 4.13\,h, \object{K2--137}\,b), which is close to its star and far from the HZ boundary, and probably near the shortest allowed value of $P_0$ (see below).
The value of 3.0\,h is also lower than the sidereal rotation periods of Jupiter and Saturn (9.925 and 10.55\,h, respectively), and might be unrealistic in the context of this work. 
Contrary to the $P_{0,1} = P_{\rm{orb}}$ scenario, all moons are stripped away (which means that $\tau_{{\rm mig},m} \, \propto \, Q\ R_p^{-5} M_p^{8/3} M_*^{-8/3}$), 
but with migration time scales greater than 0.8\,Gyr for 26 planet--moon systems, including the four systems in Table~\ref{table:Migration}.
Of them, 15 planet--moon systems have $\tau_{\rm mig} > t_{\rm H}$.
They are 
\object{GJ~96}\,b \citep{Hob18},
\object{CD--23~1056}\,b \citep{Sta17},
\object{LP~834--042}\,b \citep{Ast17b},
\object{LP~834--042}\,d \citep{Ast17b},
\object{Kapteyn's~star}\,b \citep{Ang14},
\object{BD--06~1339}\,c \citep{LoC13},
\object{PM~J11293--0127}\,d \citep{Alm15,Sin16},
\object{Ross~1003}\,b \citep{Tri18},
\object{HD~147379}\,b \citep{Rei18},
\object{HD~156384~C}\,e \citep{Ang13},
\object{CD--44~11909}\,c \citep{Tuo14},
\object{V1428~Aql}\,b \citep{Kam18},
\object{Kepler--186}\,f \citep{Tor15}, and
\object{IL~Aqr}\,b and c \citep{Tri18}.
However, except for nine cases,
\object{GJ~96}\,b,
\object{CD--23~1056}\,b,
\object{BD--06~1339}\,c,
\object{Ross~1003}\,b,
\object{HD~147379}\,b,
\object{V1428~Aql}\,b,
\object{Kepler--186}\,f, and
\object{IL~Aqr}\,b and c,
all tidal locking times are shorter than 0.8\,Gyr ($t_{\rm lock} \ll \tau_{{\rm mig},m}$), which actually transforms this scenario into the previous one, $P_{0,1} = P_{\rm{orb}}$.
For the four planets listed in Table \ref{table:Migration}, the migration time scales remain unscathed, as $a_{m} > a_{\mathrm{crit},m}$. 

\begin{figure}
\centering
\includegraphics[width=\hsize]{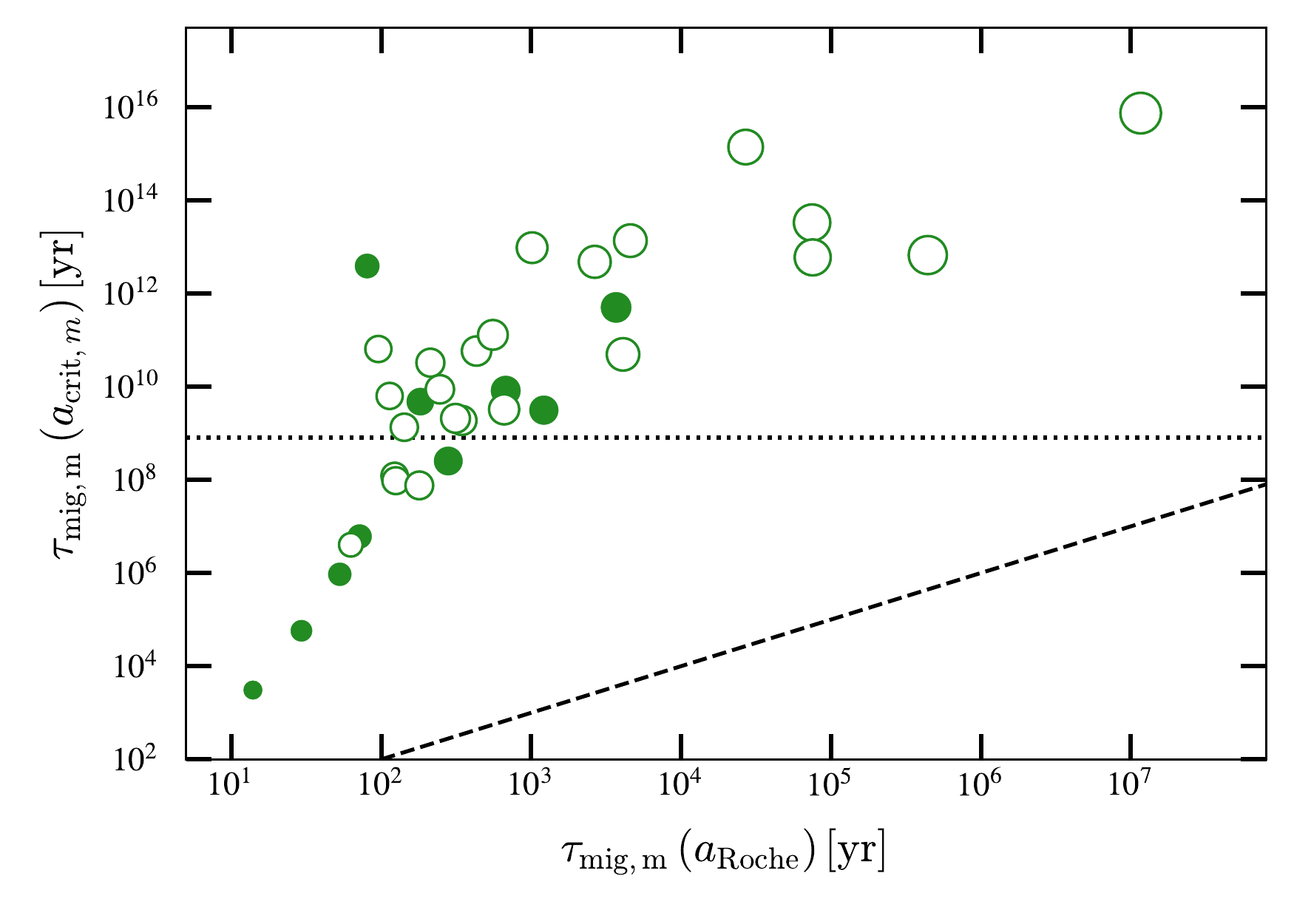}

\caption{Maximum vs. minimum allowed values of exomoon migration time scales for the potentially habitable exoplanets in our sample. The points are depicted with sizes proportional to the exoplanets' masses in logarithmic scale. The black, dashed line corresponds to the most extreme case, $a_{\mathrm{crit},m} = a_{\rm{Roche}}$. The black, dotted, horizontal line shows the 0.8\,Gyr boundary mentioned in the text (the Earth's Late Heavy Bombardment).}
\label{fig:Mig_range}
\end{figure}

To better understand the values of the exomoon migration time scales, we calculated the maximum and minimum allowed $\tau_{{\rm mig},m}$, i.e., we substituted $a_{m} = a_{\mathrm{crit},m}$ and $a_{m} = a_{\rm{Roche}}$ in Eq. \ref{eq:taumig}, respectively. Figure \ref{fig:Mig_range} shows the range of $\tau_{{\rm mig},m}$ for the potentially habitable exoplanets in our sample. In principle, all the 24 exoplanets above the Earth's Late Heavy Bombardment line might host exomoons. However, the parameter space is only broad enough for the five exoplanets on the upper right side of Figure \ref{fig:Mig_range}, which are the four planets listed in Table \ref{table:Migration} and \object{BD--06~1339}\,c. We compared our migration timescales for these five most extreme cases with the values resulting from applying the numerical technique of \citet{Pir18a} by varying the planet--moon separation over an allowable range and the spin of the planet, and found that our values are slightly longer than, but 
comparable to, these. We kept our values for consistency.

%
%

\begin{table*} 
\caption{Potentially habitable exoplanets that can host exomoons.}
\label{table:Migration}
\centering
\begin{tabular}{llccccccccccc}
\hline
\hline
\noalign{\smallskip}
Planet & GJ & $a$ & $P_{\rm{orb}}$ & $M_{*}$ & $M_{\rm{p}}$ & $R_{\rm{p}}$ & $a_{m}$ & $a_{\rm{Roche}}$ & $a_{\mathrm{crit},m}$ & ${\tau_{\mathrm{mig},m}}$ & $P_{\mathrm{orb},m}$  & $Q_{\rm tidal}$ \\ 
\noalign{\smallskip}
 & & [au] & [days] & [$M{_\odot}$] & [$M{_\oplus}$] & [$R{_\oplus}$] & $[a_{\rm{M}}]$ & $[a_{\rm{M}}]$ & $[a_{\rm{M}}]$ & [Gyr] & [days] & [TW]\\ 
\noalign{\smallskip}
\hline
\noalign{\smallskip}
CD-23 1056 b & \ldots & 0.25 & 53.435 & 0.62 & $>\ $114.0 & ($>\ $13.2) & 2.72 & 0.23 & 2.72 & $> t_{\rm{H}}$ & $\leq$ 11.0 & ${\lesssim} \, 7$ \\
Ross 1003 b & 1148 & 0.17 & 41.38 & 0.35 & $>\ $96.7 & ($>\ $12.0) & 2.08 & 0.21 & 2.08 & $> t_{\rm{H}}$ & $\leq$ 8.0 & ${\lesssim} \, 32$\\
IL Aqr b & 876 & 0.21 & 61.082 & 0.34 & $>\ $760.9 & ($>\ $13.3) & 5.36 & 0.43 & 5.36 & $> t_{\rm{H}}$ & $\leq$ 11.8 & ${\lesssim} \, 5$\\
IL Aqr c & 876 & 0.13 & 30.126 & 0.34 & $>\ $241.5 & ($>\ $14.0) & 2.29 & 0.29 & 2.29 & $> t_{\rm{H}}$ & $\leq$ 5.9 & ${\lesssim} \, 126$\\
\noalign{\smallskip}
\hline
\end{tabular}
\end{table*}
%

%

\subsection{Potentially habitable exomoons} \label{subsec:HabMoons}


As presented in Section~\ref{subsec:Migration}, only four potentially habitable exoplanets might be able to host exomoons for time scales above 0.8\,Gyr under the $P_0 = P_{\rm orb}$ scenario: CD--23~1056\,b (a giant exoplanet around an active, metal-rich, M0.0\,V star in Eridanus -- \citealt{For11}), IL~Aqr\,b,c (two giant exoplanets around a nearby M4.0\,V star with a four-planet system --  \citealt{Del98,Mar98,Mar01,Tri18}), and Ross~1003\,b (an eccentric, Saturn-mass planet around a nearby M4.0\,V star with a multiplanetary system\footnote{In the last stages of revision of this paper, we became aware of an independent study of the exomoon hypothesis around Ross 1003, by T. Trifonov et al. (in preparation).} -- \citealt{Hag10}, \citealt{Tri18}).
The four planets are relatively large and have minimum masses between 0.30 and 2.39\,$M_{\rm Jup}$ (i.e., between 97 and 761\,$M_\oplus$).

We analyzed the effects of tidal heating $Q_{\rm tidal}$ \citep[e.g.][]{Bar08,Jac08a,Jac08b,Jac08c} on the four exomoons to assess if their mantles would melt, which, for a rocky body, occurs if $Q_{\rm tidal} \gtrsim 100 \, \rm{TW}$ \citep{Moo07,Lai09,Vee12,Dri15}. \citet{BarH13} defined four categories of Earth-mass exoplanets in the HZ depending on the tidal heating that they experience: ``Earth twins'' ($Q_{\rm tidal} < 20 \, \rm{TW}$), ``Tidal Earths'' ($20 \, \rm{TW} < Q_{\rm tidal} < 1020\, \rm{TW}$), ``Super-Ios'' ($1020 \, \rm{TW} < Q_{\rm tidal} < 134000\, \rm{TW}$), and ``Tidal Venuses'' ($Q_{\rm tidal} > 134000 \, \rm{TW}$). 
We estimated $Q_{\rm tidal}$ for each exoplanet--exomoon system by means of the publicly available \texttt{VPLanet}\footnote{\tt \url{https://github.com/VirtualPlanetaryLaboratory/vplanet}} code \citep{Bar16,Bar18,Bar19}, modifying the example \texttt{IoHeat}. 
So far, we had assumed zero eccentricity and obliquity, as in \citet{Pir18a}, in which case tidal heating can only come from rotation, assuming that it is not synchronous. 
Hence, we relaxed this condition and set a small value for the eccentricity $e = 0.01$. We also assumed that the exomoon is tidally locked to the exoplanet. 
As $Q_{\rm tidal}$ is directly proportional to the moon's size, we took a Ganymede-like exomoon to estimate an upper limit for the tidal heating (Ganymede is the biggest solar system moon)\footnote{$M_{\rm Gan}$ = 1.48 $\times$ 10$^{23}$\,kg; $R_{\rm Gan}$ = 2.6341 $\times$ 10$^{6}$\,m}. 
We set the exomoon's semimajor axis $a_{m}$ to the mean distance between the Roche limit and the Hill radius from Table~ \ref{table:Migration}. 
We let each system evolve for $t = 10^{4}$ years and found that only a potential exomoon around IL~Aqr c might reach the mantle melting threshold. 
We report the upper limit for $Q_{\rm tidal}$ in Table~\ref{table:Migration}.

Some of these moons might be detected with current and near-future technology.
The detectability of moons by the transit method (or, more likely, photocentric transit timing variation) has been extensively and intensively discussed in the literature \citep{Sar99,Bar02,Sza06,Pon07,Kip09a,Kip14,Hip15,Sim15,Per18}.
The ESA space mission \citep[\textit{CHEOPS};][]{For14} might
discover moons with the transit method around the brightest M dwarfs, while, if the moons were massive and dense enough, ESPRESSO at the Very Large Telescope \citep{Pep10} could also detect the modification of the amplitude and signature of the Rossiter--McLaughlin effect as in \citet{Zhu12}. 
Unfortunately, none of the four planets in the stars' HZ with long moon migration times transits.
If the NASA space mission {\em TESS}, currently in operation \citep{Ric14}, discovered additional M-dwarf planets in this class, but transiting, they should be a high priority target for ESPRESSO or any other ultra-stable spectrograph with cm\,s$^{-1}$ accuracy at a large telescope.

Regarding the potential habitability of these Moon-mass exomoons, they
may not retain an atmosphere at 300 K due to their low gravities. As
Titan is less than twice as massive ($1.83 \, M\rm{_{M}}$), increasing the exomoon
mass to Titan's would not significantly affect our results; the
condition $a_{m} > a_{\mathrm{crit},m}$ for the four systems in Table
\ref{table:Migration} would still be met. The minimum mass to hold onto
an atmosphere in the habitable zone of a star is not known, but is
probably close to Mars' \citep[$8.71 \, M\rm{_{M}}$,][]{Pol87,Sel07}. If
we increased the masses of our putative exomoons to Mars' mass, then, for
the four systems in Table \ref{table:Migration},
${\tau_{\mathrm{mig},m}}$ would be of the order of 5 million years
for CD-23 1056\,b and Ross 1003\,b, and would lie above the Hubble time
for IL Aqr\,b anc c. These four systems would remain near the upper
right corner of Figure \ref{fig:Mig_range}, indicating that their
${\tau_{\mathrm{mig},m}}$ values are sufficiently long to still permit
a long lifetime.

\section{Conclusions} \label{sec:Conclusions}

First of all, we compiled orbital, astrometric, photometric, and basic astrophysical parameters of a comprehensive list of 205 exoplanets around 109 M dwarfs discovered with the radial velocity or transit methods.
We calculated the most probable masses and radii of 192 planets using the models of \citet{Che17}, and assumed literature values for the other 13 planets with both radial velocity and transit measurements.
For all 109 host stars, we derived luminosities, effective temperatures, and masses from public photometric catalogs, BT-Settl CIFIST models of the Lyon group, the Virtual Observatory Spectral energy distribution Analyzer, a near-infrared absolute magnitude-mass relation, and, except for only two cases, {\em Gaia} parallaxes.

With the available data, for every star we outlined the inner and outer HZ boundaries from a one-dimensional climate model.
There are 33 known planets that orbit within the HZ limits of M dwarfs.
Despite being tidally locked to their (active) host stars in most cases, it may still be possible for these planets to retain the conditions for habitability.
In this scenario, even if most of the planets in the HZ might not host liquid water, because they are icy Neptunians, some of their hypothetical exomoons could instead.

For each of the 33 planets, we modeled noneccentric, nonobliquous moons, and computed moon migration timescales for two different scenarios: 
strip-away from the planet, 
and fall-back onto the planet.
We also considered two extreme cases for the initial planetary spin: a maximum value corresponding to the orbital period (i.e. the planet is tidally locked), and a minimum value of 3\,h. 
In the $P_{0} = P_{\rm orb}$ scenario, all hypothetical moons fall back onto their planets after short time scales, except for four planets whose migration time scales are longer than the Hubble time (Ross~1003\,b, IL~Aqr\,b and~c, and CD--23~1056\,b). 
Only for these cases a hypothetical moon could be stable for time scales longer than protoplanetary disk dissipation times.
We also explored the effects of tidal heating in the exomoons, in particular, if their mantles would partially melt, similar to Io in the solar system. All our code is publicly available in a GitHub repository\footnote{\tt \url{https://github.com/hector-mr/Exomoons_HZ_Mdwarfs/}}.

Our planet compilation dates back to 2019 March 19.
As a future work, a new M-dwarf exoplanet list should be updated with recent discoveries, such as Teegarden's~b and~c \citep{Zec19} and, especially, the transiting warm Earth-mass planet GJ~357~b \citep{Luq19}.
Besides, the stellar mass determination could be improved with the new luminosity--radius--mass relation of \citet{Sch19} or, more interestingly, for studying the long-term stability of hypothetical moons, direct $N$-body analyses could be carried out.

\acknowledgements H.M.-R. acknowledges support from a PITT PACC, a Zaccheus Daniel and a Kenneth P. Dietrich School of Arts \& Sciences Predoctoral Fellowship from the Department of Physics and Astronomy at the University of Pittsburgh. 
J.A.C. and C.C. acknowledge financial support from the Agencia Estatal de Investigaci\'on of the Ministerio de Ciencia, Innovaci\'on y Universidades and the European FEDER/ERF funds through projects AYA2016-79425-C3-2-P,BES-2017-080769, and MDM-2017-0737. R.B. was supported by the NASA Virtual Planetary Laboratory Team through Grant No. 80NSSC18K0829. This work also benefited from participation in the NASA Nexus for Exoplanet Systems Science research coordination network. 
This research has made use of 
NASA's Astrophysics Data System Bibliographic Services,
the VizieR catalog access tool,
VOSA, which is developed under the Spanish Virtual Observatory project supported from the Spanish MINECO through grant AYA2017-84089,
and the NASA Exoplanet Archive, which is operated by the California Institute of Technology, under contract with the National Aeronautics and Space Administration under the Exoplanet Exploration Program. 

\software{
\texttt{SciPy} \citep{Vir19},
\texttt{Matplotlib} \citep{Hun07},
\texttt{IPython} \citep{PeG07},
\texttt{Numpy} \citep{vaW11},
\texttt{Astropy} \citep{Astro13,Astro18},
\texttt{EqTide} \citep{Bar16,Bar17},
\texttt{VPLanet} \citep{Bar16,Bar18,Bar19}
.}


\bibliographystyle{aasjournal}
\bibliography{references} 

\vspace{-15cm}

\appendix
\setcounter{table}{0}
\renewcommand{\thetable}{A\arabic{table}}

\section{Long tables}


Table \ref{table:stars} reproduces the basic properties of planet-host M dwarfs. Table \ref{table:planets} reproduces the basic properties of M-dwarf hosted exoplanets.
\setlength{\tabcolsep}{2.5pt} 

\startlongtable
\begin{deluxetable*}{lcccccccc}
\tablecaption{Basic properties of planet-host M dwarfs.\label{table:stars}}
\tablewidth{900pt}
\tabletypesize{\normalsize}
\tablehead{
Name & Alternative  &  $\alpha$\tablenotemark{a}  & $\delta$\tablenotemark{a}  & $d$\tablenotemark{b} & Sp. & $T_{\rm eff}$\tablenotemark{c} & $L$\tablenotemark{c} & $M_{*}$\tablenotemark{d} \\
 & Names & & & [pc] & type  & [K] & [10 $^{-4}\,L_{\odot}$]  & [$M_{\odot}$]  \\
} 
\startdata
GX And  &   GJ 15A  &   00:18:27.04 &   +44:01:29.0 &   3.563 $  \pm  $ 0.001   &   M1.0 V  &   3600    &   253.6 $ \pm $ 4.0   &   0.40    \\
K2-149  &   EPIC 220522664  &   00:39:17.22 &   +07:16:37.2 &   124.1 $  \pm  $ 0.5 &   M1 V    &   3700    &   677 $ \pm $ 17  &   0.59    \\
CD-44 170   &   GJ 27.1 &   00:39:59.52 &   --44:15:15.0    &   23.62 $  \pm  $ 0.03    &   M0.5 V  &   3700    &   459.7 $ \pm $ 8.7   &   0.52    \\
LHS 1140    &   GJ 3053 &   00:44:59.67 &   --15:16:26.8    &   14.99 $  \pm  $ 0.02    &   M4.5 V  &   3100    &   44.25 $ \pm $ 0.55  &   0.18    \\
2MASS J01021226-6145216 & HATS-71 & 01:02:12.45 & --61:45:22.1 &  140.79 $  \pm  $ 0.86  & M3.0 V & 3300 & 251.9 $  \pm  $ 5.5  & 0.45 \\
BD+61 195 & GJ 49 & 01:02:40.50 & +62:20:43.6 &  9.856 $  \pm  $ 0.003 & M1.5 V & 3600 & 452.4 $  \pm  $ 9.5 & 0.54 \\
YZ Cet  &   GJ 54.1 &   01:12:31.94 &   --16:59:46.5    &   3.712 $  \pm  $ 0.001   &   M4.5 V  &   3100    &   22.00 $ \pm $ 0.35  &   0.14    \\
K2-150  &   EPIC 220598331  &   01:13:22.65 &   +08:59:14.8 &   102.8 $  \pm  $ 0.8 &   M2.5 V  &   3400    &   224.3 $ \pm $ 5.5   &   0.41    \\
\enddata
\tablenotetext{$a$}{
All equatorial coordinates are given in the epoch of {\it Gaia} DR2 (J2015.5) \citep{Gaia16}, except for Luyten's star and Lalande 21185, with equatorial coordinates given in the epoch of 2MASS (J2000) \citep{Skr06}.
}
\tablenotetext{$b$}{
All distances calculated using parallaxes from \textit{Gaia} DR2 \citep{Gaia16,Gaia18}, except for Luyten's star and Lalande 21185, calculated using parallaxes from \textit{Hipparcos} \citep{van07}.
}
\tablenotetext{$c$}{
$L$ and $T_{\rm eff}$ computed with VOSA \citep{Bay08} from our compiled photometry. The uncertainties in the effective temperatures are $\delta T_{\rm eff} = 50\,K$ in all cases.
}
\tablenotetext{$d$}{
Stellar masses calculated from $K_s$ apparent magnitudes, tabulated distances, and the relations of \citet{Man19}, as in \citet{Sch19}.
}
\tablenotetext{$e$}{
Spectral type of LHS 3844 estimated from its photometric data.
}

\end{deluxetable*}

\clearpage
\begin{turnpage}
\setlength{\tabcolsep}{3pt}
\begin{longrotatetable}
\begin{deluxetable*}{llcccccccc}
\tablecaption{Basic properties of M-dwarf hosted exoplanets.\label{table:planets}}
\tablewidth{900pt}
\tabletypesize{\scriptsize}
\tablehead{
Name & Planet & $a$ & $P_{\rm{orb}}$ & $M_{\rm{p}}$\tablenotemark{a} & $R_{\rm{p}}$\tablenotemark{a} & References\tablenotemark{b} & $S_{\rm eff}$ & $T_{\rm eq, p}$ & Potentially \\
\noalign{\smallskip}
    & & [au] & [days] & [$M{_\oplus}$] & [$R_{\oplus}$] & & [$S_{\odot}$] & [K] & HZ  \\
} 
\startdata
GX And	&	b	&	$0.072^{+0.003}_{-0.004}$	&	$11.4407^{+0.0017}_{-0.0016}$	&	$>3.03^{+0.46}_{-0.44}$	&	($ > \ $1.53)	&	Pin18	&	4.89 $  \pm  $ 0.10	&	413.9 $  \pm  $ 8.7	&    \\
K2-149	&	b	&	$0.083^{+0.027}_{-0.027}$	&	$11.332^{+0.0013}_{-0.0013}$	&	({3.85})	&	$1.64^{+0.20}_{-0.18}$	&	Hir18	&	9.82 $  \pm  $ 0.25	&	493 $  \pm  $ 13	&    \\
CD-44 170	&	b	&	$0.101^{+0.009}_{-0.013}$	&	$15.8190^{+0.0049}_{-0.0026}$	&	$>13^{+4.1}_{-6.6}$	&	($ > \ ${3.63})	&	Tuo14	&	4.51 $  \pm  $ 0.35	&	406 $  \pm  $ 32	&    \\ 
LHS 1140	&	b	&	$0.0875^{+0.0041}_{-0.0041}$	&	$24.73712^{+0.00025}_{-0.00025}$	&	$6.65^{+1.82}_{-1.82}$	&	$1.43^{+0.10}_{-0.10}$	&	Dit17	&	0.5780 $  \pm  $ 0.0072	&	242.7 $  \pm  $ 3.0	&  Yes  \\ 
	&	c	&	$0.02675^{+0.00070}_{-0.00070}$	&	$3.777931^{+0.000006}_{-0.000006}$	&	$1.81^{+0.39}_{-0.39}$	&	$1.282^{+0.024}_{-0.024}$	&	Men19	&	6.184 $  \pm  $ 0.077	&	438.9 $  \pm  $ 5.4	&    \\ 
2MASS J01021226-6145216  	&	b	&	$ 0.03946^{+ 0.00095}_{-0.00168 }$	&	$3.7955213 ^{+0.0000011}_{-0.0000011 }$	&	({1750})	&	$12.106 ^{+ 0.179}_{-0.179}$	&	 Bak18	&	  16.18 $  \pm  $ 1.14 	&	 558  $  \pm  $ 39 	& \\ 
BD+61 195	&	b	&	$0.0905 ^{+ 0.0011}_{-0.0011 }$	&	$ 13.8508^{+ 0.0053}_{-0.0051 }$	&	$>5.63 ^{+ 0.67}_{-0.68 }$	&	($ > \ ${2.25})	&	 Per19	&	 5.52 $  \pm  $ 0.12 	&	426.7  $  \pm  $ 9.0 	&    \\ 
YZ Cet	&	b	&	$0.01557^{+0.00052}_{-0.00052}$	&	$1.96876^{+0.00021}_{-0.00021}$	&	$>0.75^{+0.13}_{-0.13}$	&	($ > \ ${0.93})	&	Ast17b	&	9.08 $  \pm  $ 0.14	&	483.1 $  \pm  $ 7.6	&    \\
\enddata
\end{deluxetable*}
\end{longrotatetable}

\end{turnpage}
\clearpage
\global\pdfpageattr\expandafter{\the\pdfpageattr/Rotate 90}


\end{document}